\documentclass[twocolumn,showpacs,showkeys]{revtex4}
\usepackage{graphicx}
\input{psfig.sty}

\newcommand{\bastar}{\begin{eqnarray*}}
\newcommand{\eastar}{\end{eqnarray*}}
\newskip\humongous \humongous=0pt plus 1000pt minus 1000pt

\newif\ifdtup

\relax
\newcommand{\be}{\begin{equation}}
\newcommand{\ee}{\end{equation}}
\newcommand{\bea}{\begin{eqnarray}}
\newcommand{\eea}{\end{eqnarray}}

\newcommand{\pro}{\partial}

\newcommand{\dfrac}{\displaystyle\frac}
\newcommand{\ba}{\begin{array}}
\newcommand{\ea}{\end{array}}

\newcommand{\nn}{\nonumber}

\begin{document}
\title{Vorticity Knot in Two-component Bose-Einstein Condensates}
\author{Y. M. Cho}
\email{ymcho@yongmin.snu.ac.kr}
\affiliation{School of Physics, College of Natural Sciences, 
Seoul National University,
Seoul 151-742, Korea  \\
and \\
C.N.Yang Institute for Theoretical Physics, State University of
New York, Stony Brook, NY 11790, USA}
\begin{abstract}
~~~~~We demonstrate the existence of the helical vortex 
solution in two-component Bose-Einstein condensates which
can be identified as a twisted vorticity flux. Based on this
we argue that the recently proposed knot in two-component Bose-Einstein 
condensates can be interpreted as a vorticity knot, a vortex
ring made of the helical vortex. This picture shows that 
the knot is made of two quantized vorticity fluxes linked together, 
whose topology $\pi_3(S^2)$ is fixed by the linking number of two 
vorticity fluxes. Due to the helical structure the knot has both
topological and dynamical stability. We estimate the energy 
of the lightest knot to be about $3\times 10^{-3}~eV$.
\end{abstract}
\maketitle

The topological objects, in particular finite energy topological
objects, have played an important role in physics \cite{abri,skyr}. 
In Bose-Einstein condensates (BEC) the best known topological 
objects are the vortices, which have been widely studied
in the literature. Theoretically these vortices have 
successfully been described by the Gross-Pitaevskii 
Lagrangian. On the other hand, the recent advent of
multi-component BEC (in particular
the spin-1/2 condensate of $^{87}{\rm Rb}$ atoms)  
has widely opened a new opportunity for
us to study novel topological objects which can not be
realized in ordinary (one-component) BEC \cite{exp1,exp2}.
This is because the multi-component BEC naturally allows 
a non-Abelian structure which accomodates a non-trivial 
topological objects, in particular a topolgical knot which is 
very similar to the knot in Skyrme theory \cite{bec1,cho01}.

Indeed recently many authors have proposed the existence of a knot
in Gross-Pitaevskii theory of two-component BEC \cite{ruo,batt,met}. 
{\it The purpose of this report is to show that this knot is
nothing but a vorticity knot which is made of
two vorticity fluxes linked together. Furthermore, we
show that the knot is topological, whose topology $\pi_3(S^2)$ 
is fixed by the Chern-Simon index of the velocity potential 
of the condensate.} To show this we first present a helical 
vortex solution in two-component BEC which is periodic in 
$z$-coordinate, and construct a helical vortex ring 
by bending it and smoothly connecting    
two periodic ends together. We show that this vortex ring 
becomes the vorticity knot whose quantum number is fixed by the 
Chern-Simon index of the velocity potential,
which describes the linking number of two vorticity fluxes.

This picture tells that the knot has both topological 
and dynamical stability. The topological stability follows from
the fact that two linked vorticity fluxes can not be disconnected 
by any smooth deformation of the field configuration. The 
dynamical stability follows from the fact that the knot necessarily 
has a net velocity flux along the knot, and thus a non-vanishing
angular momentum around the knot. This creates a repulsive 
stablizing force against the collapse of the knot.
This provides the dynamical stability of the knot.
 
The knot that we discuss here are very similar to the knot in 
Skyrme theory \cite{cho01,fadd,sky1}. 
Just as the knot in Skyrme theory is a vortex ring
made of the helical magnetic vortex,
our knot here is a vortex ring made of the helical vorticity 
vortex. So it is crucial that we have the helical vortex
to demonstrate the existence of the vorticity knot in 
two-component BEC.
 
To construct the desired vortex solution 
let the two-component BEC be a complex doublet $\phi=(\phi_1,\phi_2)$,
and consider the Lagrangian 
\bea
&{\cal L} = i \dfrac{\hbar}{2} \phi^\dag \partial_t \phi
- \dfrac {\hbar^2}{2M} |\partial_i \phi|^2 
+ \mu_1 \phi_1^\dag \phi_1 + \mu_2 \phi_2^\dag \phi_2 \nn\\
&- \dfrac {\lambda_{11}}{2} (\phi_1^\dag \phi_1)^2
- \lambda_{12} (\phi_1^\dag \phi_1)(\phi_2^\dag \phi_2)
- \dfrac {\lambda_{22}}{2} (\phi_2^\dag \phi_2)^2,
\label{gplag1}
\eea
where $\mu_i$ are the quadratic coupling constants 
and $\lambda_{ij}$ are the quartic coupling constants
which are determined by the scattering
lengths $a_{ij}$
\bea
\lambda_{ij}=\dfrac{4\pi {\hbar}^2}{M} a_{ij}.
\eea
This is an obvious generalization of one-component Gross-Pitaevskii
Lagrangian to the two-component BEC. Notice that here we have
neglected the trapping potential, because we are assuming that
the range of the trapping potential is much larger than
the size of tpological objects we are interested in.

Clearly the Lagrangian has a global
$U(1)\times U(1)$ symmetry. But one could simplify it
because experimentally
the scattering lengths often have almost the same value.
For example, for the spin $1/2$ condensate of $^{87}{\rm Rb}$
atoms, all $a_{ij}$ are about $5.5~nm$ and differ by only about
$3~\%$ or so \cite{exp1,exp2}. In this case one may safely assume
$\lambda_{11} \simeq \lambda_{12} \simeq \lambda_{22} 
\simeq \bar \lambda$. 
With this the Lagrangian is written as
\bea
&{\cal L} = i \dfrac {\hbar}{2} \phi^\dag \partial_t \phi
- \dfrac {\hbar^2}{2M} |\partial_i \phi|^2 
-\dfrac{\bar \lambda}{2} \big(\phi^\dag \phi
-\dfrac{\mu}{\bar \lambda} \big)^2 \nn\\
&- \delta \mu \phi_2^\dag \phi_2,
\label{gplag2}
\eea
where $\mu=\mu_1$ and $\delta \mu = \mu_1-\mu_2$.
Notice that the Lagrangian has a global $U(2)$ symmetry
when $\delta \mu=0$. So the $\delta \mu$ interaction is the symmetry
breaking term which breaks the global $U(2)$ symmetry to
$U(1)\times U(1)$. This means that even when $\delta \mu \ne 0$
the Lagrangian has an approximate $U(2)$ symmetry.
Physically $\delta \mu$ can be viewed to represent
the difference of the chemical potentials between $\phi_1$
and $\phi_2$, so that it does not vanish 
when the chemical potentials are different.

With
\bea
\phi = \dfrac {1}{\sqrt 2} \rho \zeta , ~~~~~(\zeta^\dag \zeta = 1)
\label{phi}
\eea
the Lagrangian (\ref{gplag2}) gives the following Hamiltonian
in the static limit (in the natural unit $c=\hbar=1$),
\bea
&{\cal H} =  \dfrac {1}{2} (\pro_i \rho)^2
+ \dfrac {1}{2} \rho^2 |\pro_i \zeta|^2
+ \dfrac{\lambda}{8} (\rho^2-\dfrac{2\mu}{\lambda})^2 \nn\\
&+ \dfrac{\delta \mu^2}{2} \rho^2 \zeta_2^*\zeta_2,
\label{gpham1}
\eea
where $\lambda=4M^2 \bar \lambda,~\delta \mu^2=2M\delta \mu,
~\rho_0^2=4\mu M/\lambda$, 
and we have normalized $\rho$ to
$(\sqrt{2M}/\hbar)\rho$. The Hamiltonian (\ref{gpham1}) 
can be expressed as
\bea
&{\cal H} = \lambda \rho_0^4~\Big\{ \dfrac {1}{2} (\hat \pro_i \hat \rho)^2
+ \dfrac {1}{2} \hat \rho^2 |\hat \pro_i \zeta|^2
+ \dfrac{1}{8} (\hat \rho^2-1)^2 \nn\\
&+ \dfrac{\delta \mu}{4\mu} \hat \rho^2 \zeta_2^*\zeta_2 \Big\},
\label{gpham2}
\eea
where $\hat \rho=\rho/\rho_0$ and 
$\hat \pro_i=\pro_i/\sqrt \lambda \rho_0$. 
This tells that the physical unit of the Hamiltonian is
$\lambda \rho_0^4$, and the physical scale $\kappa$ of the coordinates
is $1/\sqrt \lambda \rho_0$. This is comparable to
the correlation length $\bar \xi=1/\sqrt {2\mu M}$.
Indeed we have $\kappa=\bar \xi/\sqrt 2$.

\begin{figure}
\includegraphics[scale=0.7]{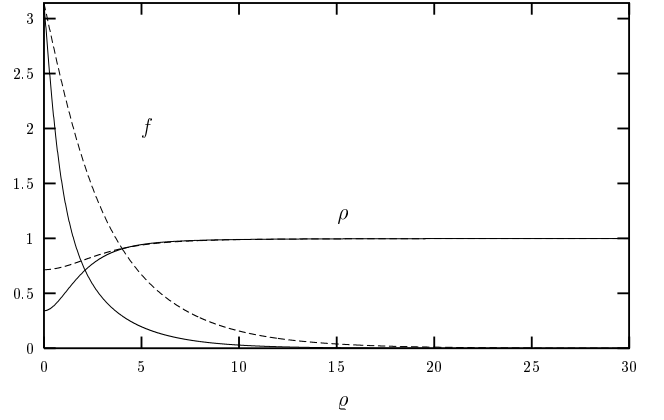}
\caption{The helical vortex in the Gross-Pitaevskii
theory of two-component BEC. Here we have put $m=1, m'=-1, n=1, n'=0, 
k=0.25 /\kappa$, and $\varrho$ is in the unit of $\kappa$.
Dashed and solid lines correspond to
$\delta \mu/\mu=0$, and $0.1$ respectively.}
\label{twobec-fig}
\end{figure}

From the Hamiltonian we have
\bea
& \pro^2 \rho - |\pro_i \zeta|^2 \rho
=\Big (\dfrac{\lambda}{2} (\rho^2-\rho_0^2)
+ \delta \mu^2 (\zeta_2^* \zeta_2) \Big) \rho, \nn\\
& \Big\{(\pro^2 - \zeta^\dag \pro^2 \zeta) + 2 \dfrac {\pro_i
\rho}{\rho}(\pro_i - \zeta^\dag \pro_i\zeta) 
+\delta \mu^2 (\zeta_2^* \zeta_2) \Big\} \zeta_1 \nn\\
&= 0, \nn\\
&\Big\{(\pro^2 - \zeta^\dag \pro^2 \zeta) + 2 \dfrac {\pro_i
\rho}{\rho}(\pro_i - \zeta^\dag \pro_i\zeta) 
-\delta \mu^2 (\zeta_1^* \zeta_1) \Big\} \zeta_2  \nn\\
&= 0.
\label{gpeq1}
\eea
To obtain the vortex solution, we choose the ansatz
\bea
&\rho= \rho(\varrho), \nn\\
&\zeta = \exp(-i\gamma)~\xi,~~~~~\gamma=n'\varphi+m'kz,  \nn\\
&\xi = \Bigg( \matrix{\cos \dfrac{f(\varrho)}{2}
\exp (-in\varphi-imkz) \cr \sin \dfrac{f(\varrho)}{2} } \Bigg).
\label{gpans}
\eea
Now, with $n'=0$ and $m'=-m$ (\ref{gpeq1}) is reduced to
\bea
&\ddot{\rho}+\dfrac{1}{\varrho}\dot{\rho}
-\bigg(\dfrac{1}{4}\dot{f}^2 + \dfrac{n^2}{\varrho^2} \nn\\
&-\big(\dfrac{n^2}{\varrho^2}
-m^2 k^2 - \delta \mu^2 \big)\sin^2{\dfrac{f}{2}}\bigg)\rho
= \dfrac{\lambda}{2} (\rho^2-\rho_0^2) \rho,\nn \\
&\ddot{f}+\bigg(\dfrac{1}{\varrho}
+2\dfrac{\dot{\rho}}{\rho}\bigg)\dot{f}
+\bigg(\dfrac{n^2}{\varrho^2}  
-m^2k^2 - \delta \mu^2\bigg)\sin{f} \nn\\
&=0.
\label{gpeq3}
\eea
So with the boundary condition
\bea
&\rho'(0)=0,~~\rho(\infty)=\rho_0,~~f(0)=\pi,~~f(\infty)=0,
\label{gpbc}
\eea
we can solve (\ref{gpeq3}).
With $m=n=1$ we obtain the twisted
vortex solution shown in Fig. \ref{twobec-fig}.

The untwisted non-Abelian vortex solution
has been discussed before \cite{met}, but the twisted
vortex solution here is new. Notice that
when $\delta \mu^2=0$, there is no untwisted vortex solution 
because in this case the vortex size become infinite. 
But remarkably the helical vortex exists even when $\delta \mu^2=0$.
This is because the twisitng reduces the size of vortex tube.

\begin{figure}
\includegraphics[scale=0.5]{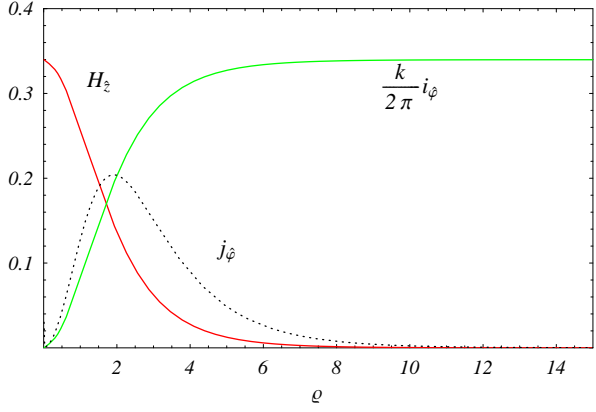}
\caption{The supercurrent $i_{\hat \varphi}$ (in one period section
in $z$-coordinate) and corresponding
magnetic field $H_{\hat z}$ circulating around the cylinder of
radius $\varrho$ of the helical vortex in two-component BEC. 
Here $m=1,m'=-1,n=1,n'=0$,
$~k=0.25/\kappa$, and $\varrho$ is in the unit of $\kappa$.
The current density $j_{\hat \varphi}$ is
represented by the dotted line.}
\label{beciphi-fig}
\end{figure}

In Skyrme theory the helical vortex is interpreted as
a twisted magnetic vortex whose flux is quantized \cite{cho01,sky1} 
Now we show that the above vortex is a twisted vorticity vortex. 
To see this notice that the non-Abelian structure of 
the vortex is represented by the doublet $\zeta$. 
Moreover, the velocity field 
of the doublet is given by \cite{bec1}
\bea
&V_\mu = i\zeta^{\dagger} \pro_\mu \zeta
=i\xi^{\dagger} \pro_\mu \xi + \pro_\mu \gamma \nn\\
&=\dfrac{1}{2}(\cos{f(\varrho)}+1)(n \pro_\mu \varphi + mk \pro_\mu z)
+ \pro_\mu \gamma,
\label{gpvel}
\eea
which generates the vorticity 
\bea
&V_{\mu\nu}= \pro_\mu V_\nu - \pro_\nu V_\mu
= i(\pro_\mu \xi^{\dagger} \pro_\nu \xi
-\pro_\nu \xi^{\dagger} \pro_\mu \xi) \nn\\
&=-\dfrac{\dot{f}}{2} \sin{f}\Big(n(\pro_\mu \varrho \pro_\nu \varphi
-\pro_\nu \varrho \pro_\mu \varphi) \nn\\
&+mk(\pro_\mu \varrho \pro_\nu z
- \pro_\nu \varrho \pro_\mu z) \Big).
\label{gpvor}
\eea
This has two vorticity fluxes, $\phi_{\hat z}$
along the $z$-axis
\bea
&\phi_{\hat z}=\dfrac{}{} \int V_{{\hat \varrho}{\hat \varphi}}
\varrho d \varrho d \varphi
= 2\pi n,
\label{gpfluxz}
\eea
and $\phi_{\hat \varphi}$ around the the $z$-axis (in one
period section from $z=0$ to $z=2\pi/k$)
\bea
&\phi_{\hat \varphi}=\dfrac{}{} \int_0^{2\pi/k}
V_{{\hat z}{\hat \varrho}} d \varrho dz
= - 2\pi m.
\label{gpfluxphi}
\eea
Obviously they are quantized. As importantly they are linked 
together, and have the linking number $mn$.

\begin{figure}
\includegraphics[scale=0.5]{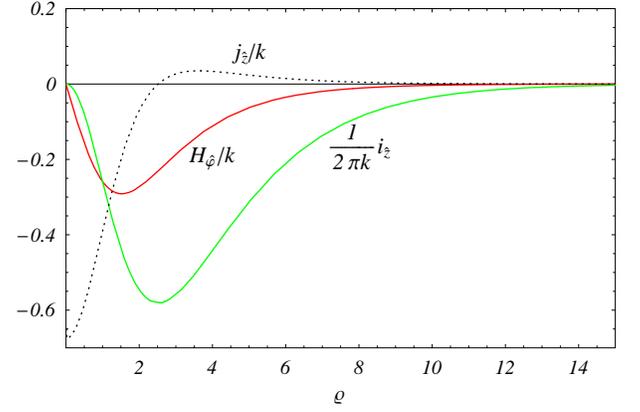}
\caption{The supercurrent $i_{\hat z}$ and corresponding
magnetic field $H_{\hat \varphi}$ flowing through the disk of
radius $\varrho$ of the helical vortex in two-component BEC. 
Here $m=1,m'=-1,n=1,n'=0$,
$k=0.25 /\kappa$, and $\varrho$ is in the unit of $\kappa$.
The current density $j_{\hat z}$
is represented by the dotted line.}
\label{beciz-fig}
\end{figure}

Furthermore, just as in Skyrme theory, these fluxes can be
viewed to originate from the helical supercurrent which confines
them with a built-in Meissner effect \cite{sky1}
\bea
&j_\mu = \pro_\nu V_{\mu\nu} \nn\\
&=\sin f \Big[n \big(\ddot f + \dfrac{\cos f}{\sin f}
\dot f^2 - \dfrac{1}{\varrho} \dot f \big) \partial_{\mu}\varphi \nn\\
&+mk \big(\ddot f + \dfrac{\cos f}{\sin f} \dot f^2
+ \dfrac{1}{\varrho} \dot f \big) \partial_{\mu}z \Big].
\label{gpsc}
\eea
This produces the supercurrents $i_{\hat\varphi}$ (in one
period section from $z=0$ to $z=2\pi/k$)
around the $z$-axis
\bea
&i_{\hat\varphi} = \dfrac{2 \pi n}{k}\dfrac{\sin{f}}{\varrho}\dot f
\Big|_{\varrho=0}^{\varrho=\infty},
\eea
and $i_{\hat z}$ along the $z$-axis
\bea
&i_{\hat z} = 2 \pi mk \varrho \dot f \sin{f}
\Big|_{\varrho=0}^{\varrho=\infty}.
\eea
The vorticity fluxes and the corresponding supercurrents
are shown in Fig. \ref{beciphi-fig} and Fig. \ref{beciz-fig}.
This is strikingly similar to what we find in the magnetic vortex 
in Skyrme theory \cite{sky1}.
This tells that the helical
vortex is nothing but the twisted vorticity flux
confined along the $z$-axis by the velocity current,
whose flux is quantized due to the topological reason.
We emphasize that this
interpretation holds even when the $\delta \mu^2$ is not zero.

We can estimate the energy of the helical vortex. 
For $^{87}{\rm Rb}$ we have
\bea
&M \simeq 8.1 \times 10^{10}~eV,
~~~~~\bar \lambda \simeq 1.68 \times 10^{-7}~(nm)^2, \nn\\
&\mu \simeq 3.3 \times 10^{-12}~eV,
~~~~~\delta \mu \simeq 0.1~\mu.
\eea
So, with $m=n=1,~m'=-1,~n'=0$ and $k=0.25/\kappa$, we find numerically 
that the energy per one periodic section 
(from $z=0$ to $z=2\pi/k$) is given by
\bea
&E \simeq 270.987~\dfrac{\rho_0^2}{\sqrt \lambda~\mu}
\simeq 1492.4~\rho_0 \nn\\
&\simeq 2.29 \times 10^{-3}~eV.
\label{gphve2}
\eea
As we will see later, the lightest knot could have an energy 
comparable to this energy.

Notice that the vorticity (\ref{gpvor}) is completely
fixed by the $CP^1$ field $\xi$, because it does not depend on
the $U(1)$ phase $\gamma$ of $\zeta$. Moreover
$\xi$ naturally defines a mapping from
the compactified $xy$-plane $S^2$ to the target space $S^2$.
This means that our vortex has exactly the same topological origin
as the baby skyrmion in Skyrme theory, but now the topological
quantum number is expressed by $\pi_2(S^2)$
of the condensate $\xi$,
\bea
&q = - \dfrac {i}{4\pi} \int \epsilon_{ij} \partial_i \xi^{\dagger}
\partial_j \xi  d^2 x = n.
\label{gpvqn}
\eea
This clarifies the topological origin of the non-Abelian vortex
in two-component BEC.

The helical vortex will become
unstable unless the periodicity condition is enforced by
hand. But just as in Skyrme theory we can make it 
a stable knot by smoothly connecting
two periodic ends. In this knot the periodicity condition is 
automatically guaranteed, and 
the very twist which causes the instability of the helical vortex
now ensures the stability of the knot. This is so
because dynamically the momentum $mk$ along the $z$-axis
created by the twist now generates a velocity current and thus
a net angular momentum which provides 
the centrifugal repulsive force
preventing the knot to collapse. 

Furthermore, this dynamical stability of the knot is now
backed up by the topological stability. This is because
mathematically the doublet $\xi$, after forming a knot,
acquires a non-trivial topology $\pi_3(S^2)$. And the
the knot quantum number is given by the Chern-Simon index of the
velocity potential,
\bea
&Q = - \dfrac {1}{4\pi^2} \int \epsilon_{ijk} \xi^{\dagger}
\partial_i \xi ( \partial_j \xi^{\dagger}
\partial_k \xi ) d^3 x \nn\\
&= \dfrac{1}{16\pi^2} \int \epsilon_{ijk} V_i V_{jk} d^3x
=mn.
\label{bkqn}
\eea
This is precisely the linking number of two vorticity fluxes.
As importantly, this is formally identical to the knot 
quantum number in Skyrme theory \cite{cho01,fadd,sky1}.
This assures the topological stability of the knot,
because two fluxes linked together can not be disconnected by any
smooth deformation of the field configuration.

We can estimate the energy of the knot, noticing
that the radius of the lowest energy vortex ring is about four times
the vortex tube size \cite{cm3}. This suggestes that
the lightest knot has the energy comparable to
the energy of the lightest helical vortex in one periodic section 
with $k\simeq 1/4\kappa$. So the lightest knot in $^{87}{\rm Rb}$ 
is expected to have the energy of the order of $3\times 10^{-3}~eV$.

The existence of a knot in Gross-Pitaevskii 
theory of two-component BEC has been proposed by several authors
\cite{ruo,batt,met}. In this paper we have clarified the physical
meaning of the knot. Just as the knot in Skyrme theory 
is a twisted magnetic flux ring,
this knot is a twisted vorticity flux ring.
It has a topological quantum number 
given by the Chern-Simon
index of the velocity potential of the condensate,
and enjoys both topological and dynamical stability.

What is remarkable is that this knot is almost identical to 
the knot in the gauge theory of two-component BEC that
we proposed recently \cite{bec1}. Both are vorticity knots
whose topology is identical. This implies that we have two 
competing theories of two-component BEC, the Gross-Pitaevskii
theory and the recently proposed gauge theory, which
can describe the knot. 

Constructing the knot might not be simple, 
but might have already been done \cite{exp2,exp3}.
Identifying it as a vorticity knot, however, may be a challenging
task. A detailed discussion on the subject will 
be published elsewhere \cite{bec3}.
                                                                               
{\bf ACKNOWLEDGEMENT}

~~~The work is supported in part by the Basic Research Program of
Korea Science and Enginnering Foundation (Grant R02-2003-000-10043-0),          and by the BK21 project of the Ministry of Education.

\end{document}